# A NON-ISOTHERMAL, NON-EQUIMOLAR TRANSIENT KINETIC MODEL FOR GAS-SOLID REACTIONS

Fabrice PATISSON[†], Magda GALANT FRANÇOIS[‡] and Denis ABLITZER[†]

[†]Laboratoire de Science et Génie des Matériaux Métalliques, Ecole des Mines, Parc de Saurupt, 54042 Nancy Cedex, France

*now (2014) with Institut Jean Lamour, Parc de Saurupt, 54011 Nancy Cedex, France,*
*E-mail : fabrice.patisson@univ-lorraine.fr*

[‡]PPGEMM, Universidade Federal do Rio Grande do Sul, Porto Alegre, Brazil

**Abstract** — A numerical model is presented, designed to simulate the kinetic and thermal behaviour of a porous pellet in which any gas-solid reaction is taking place. Its novelty consists in the fact that it can deal with reactions whether they are exothermic or endothermic, whether they are equimolar or not, whether they are reversible or irreversible, and further reactions in the transient regime and even the possible presence of inert gases and solids can be treated. The numerical scheme is based on the finite volume method in an implicit formulation, with a specific treatment of the thermal source term for strongly exothermic reactions. The model was validated by comparison with analytical and numerical solutions from the literature and was used to simulate the exothermic reaction involved in the oxidation of zinc sulphide.

# 1. INTRODUCTION

The diversity and economic importance of solid transformation processes controlled by the kinetics of gas-solid reactions explain the large number of models describing these reactions that have been published in the literature. The model presented here has been developed in order to simulate any heterogeneous exo- or endothermic reaction in the transient regime

$$a\, A_{(g)} + b\, B_{(s)} \leftrightarrows p\, P_{(g)} + q\, Q_{(s)} \qquad (1)$$

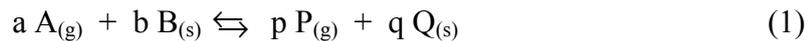

taking place within a porous pellet immersed in a gas. Although reactions of this type are commonplace, as will be shown below, in anisothermal conditions and in the transient regime, they cannot be treated with the models presently available. This observation led us to undertake the present work.

Since the basic Shrinking Core Model (SCM) and Uniform Conversion Model (UCM) (Levenspiel, 1979), kinetic models have become progressively more sophisticated, in order to more closely describe the complex simultaneous phenomena involved in gas-solid reactions, including the chemical reaction itself, diffusional transport on different scales in the porous solid, external mass transfer, heat transfer, etc. A good way of classifying the different models in the literature and of situating the present one is to distinguish them according to the necessary assumptions on which they are based.

The pseudo-steady state approximation, which consists in neglecting the accumulation in the gaseous phase, is generally recognized to be valid for gas-solid reactions (Bischoff, 1963; Bhatia, 1985) and is therefore integrated in the majority of models. However, in the case of anisothermal models, the term expressing the heat accumulation in the porous solid cannot be neglected (Wen and Wang, 1970).

The equimolar counter-diffusion assumption, which implies that the reactive gas flux and product gas flux are equal and opposite, is not normally valid, except under the pseudo-steady state approximation when the stoichiometric coefficients of each of the gases are equal (a = p). In spite of this, this assumption is most frequently made, merely because it greatly simplifies the equations for diffusional transport in the pellet. Nevertheless, Sohn and Sohn (1980), Ulrichson and Mahoney (1980) and Sohn and Bascur (1982) have clearly established the influence of the total flux induced by the non-equimolarity whenever the diffusional resistances are not negligible.

The assumption of isothermal conditions, which is also usually made in the majority of gas-solid reaction models, must be questioned whenever the reaction is exo- or endothermic or when a temperature variation is imposed in the external gas. Two types of anisothermal behaviour are then possible: the temperature of the pellet, which changes with time, can either be uniform at all points or internal temperature gradients may occur. Anderson (1963) proposed a criterion for predicting the possibility of such gradients, by comparing the heat evolved or consumed by the reaction to that which can be conducted through the pellet.

Among the studies devoted to anisothermal reactions, the earliest work (Cannon and Denbigh, 1957; Shen and Smith, 1965; Ishida and Wen, 1968b), using the SCM in the steady state regime, mainly discussed a specific feature of exothermic gas-solid reactions, namely the possibility of multiple stationary states. It was Beveridge and Goldie (1968) and Wen and Wang (1970) who first showed, as already pointed out, the importance of the transient term in the heat balance, with the result that certain theoretically possible steady states are never attained in practice. If the anisothermal models that ignore this accumulation term are excluded, together with those based on the simple SCM, those which neglect internal temperature gradients and those which do not take into account the heat of reaction, there remains only the work of Ishida et al. (1971), Sampath et al. (1975) and Hindmarsh and Johnson (1988; 1991). Ishida et al. mainly investigated thermal instability phenomena, leading to ignition or extinction, and described their experimental manifestations. Sampath et al. considered an irreversible first order equimolar reaction and proposed a fairly powerful anisothermal transient grain model including external mass and heat transfer resistances. They clearly illustrate the influence of different parameters on the transient and anisothermal behaviour of the pellet. Hindmarsh and Johnson developed by steps what we consider to be the most complete modelling of anisothermal, non-equimolar, transient gas-solid reactions, first of all for a binary gas (1988), then for a multicomponent gas with several reactions (1991). However, their work places a strong emphasis on the numerical issues, mainly because they used multipurpose differential-algebraic equation solvers which necessitate thorough conditioning of the system.

The last assumption to be examined is that concerning the structure of the pellet. Depending on the description of this structure, a distinction can be made between grain models, pore models, and pseudo-homogeneous models.

The original Grain Model, presented in detail by Szekely et al. (1976), assumes the porous pellet to be made up of small grains all of the same size and initially non porous, which react individually according to the SCM. The principal mechanisms involved in mass transfer, namely chemical reaction per se, diffusion in the layer of solid produced, diffusion in the intergranular pores, and transfer through the external boundary layer, are taken into account. Because it represents a significant advance compared to basic models such as the SCM, and since its description of the structure corresponds effectively to that of numerous porous solids, the Grain Model has been extensively employed and has proved quite successful. However, the simplifying assumptions it makes (pseudo-steady state approximation, isothermal conditions, equimolar reaction of first order with respect to A, no variation in structure or porosity) cannot be overlooked and led various authors to develop extensions to the original model. The pseudo-steady state approximation and the isothermicity and equimolarity assumptions have already been discussed. The variation of the porosity was often introduced, particularly by Ranade and Harrison (1979) and by Georgakis et al. (1979) with their Changing Grain Size Model. Kimura et al. (1981c) and Heesink et al. (1993) took into account a distribution of grain sizes, together with a variation of grain size in the model of Heesink et al. Alvfors and Svedberg (1988) presented the Partially Sintered Spheres Model, while Dam-Johansen et al. (1991) developed a three-scale model, with grains composed of micrograins.

However, authors who dealt with a large variation in porosity, with either formation and growth of pores, as in the case of char gasification, or closure and obstruction of pores, as during the sulphation of lime, have tended to prefer pore models. These models consider the solid phase to be continuous and describe the geometrical variation of the pores due to the reaction. Like the grain models, the pore models have become increasingly sophisticated since the early publications (Ramachandran and Smith, 1977; Bhatia and Perlmutter, 1980, 1981; Gavalas, 1980, 1981). Sahimi et al. (1990) reviewed the corresponding literature, extending it to include statistical models of pore networks.

Finally, it should be noted that the structural models, although apparently different, are not in fact totally independent. They all attempt to describe the same physical phenomena related to the reaction, and the description of the structure in terms of either grains or pores leads essentially to a specific calculation of the reaction surface, which changes in the course of the conversion. As shown by Bhatia and Perlmutter (1983), it is virtually possible to go from one model to the other by modifying their parameters. It is also possible to interpret the same experimental results with a grain model, a pore model, or even a pseudo-homogeneous model (Miura, 1989). The models of the pseudo-homogeneous type (Kimura et al., 1981a, 1981b, 1982; Grzesik, 1991) do not assume any particular structure and consider the gas and solid as a continuum. As will be shown, they can be tantamount to certain grain and pore models, depending on the expression of the volume reaction rate.

This introduction thus shows that the transient, anisothermal and non-equimolar model presented below has no equivalent in the literature. Only Hindmarsh and Johnson (1988; 1991) have considered the same conditions simultaneously, but their approach is more difficult to handle than the present model.

## 2. MATHEMATICAL MODEL

### 2.1. Principle

The present model describes the kinetic and thermal behaviour of a porous pellet in which a gas-solid reaction of the (eq. 1) type takes place. It is a model of the homogeneous type which treats the porous solid as a continuum. It takes account of external transfer of the gaseous species to the surface of the pellet, diffusional mass transport in the gas within the pores, the heterogeneous reaction, the generation or consumption of heat by the reaction, heat transfer by effective conduction, and heat exchange with the external environment.

Some of the classical assumptions discussed in the introduction are removed in the present model:

- the reaction can be reversible, non-equimolar ($a \neq p$), exo- or endothermic,
- allowance is made for the possible presence of inert gaseous (I) or solid (J) species,

- 6 -

- the regime is transient and not pseudo-steady state,
- various reaction rate kinetics can be considered,
- various diffusion regimes are possible.

Conversely, the following assumptions are maintained:

- the pellet is spherical and its diameter remains constant,
- the total gas pressure is constant,
- the porosity and the tortuosity are not modified by the progression of the reaction.

The model is modular in conception and in future versions could be made more general by eliminating some of the previous assumptions. In particular, the last assumption is certainly the most restrictive. As already mentioned in the introduction, sophisticated descriptions of the variation in porosity are available in some models from the literature. This aspect has not yet been introduced in the present model, because we gave priority to handling anisothermal conditions.

*2.2. Equations*

The model is one-dimensional and transient. The variables and parameters are therefore functions of r, the radial position in the pellet, and of the time t. The notation is listed at the end of the text. Throughout what follows, the variables $x_A$, $x_P$, $N_t$, $f_B$, $c_Q$, $c_J$ and T will be termed principal variables, and are calculated respectively from each of the conservation equations given below.

2.2.1. Conservation equations

Molar balance for the gaseous reactant A

$$\mathrm{div}\mathbf{N}_A + \frac{\partial}{\partial t}\left(\varepsilon\, c_t\, x_A\right) = -a\, v \tag{2}$$

Molar balance for the gaseous product P

$$\mathrm{div}\mathbf{N}_P + \frac{\partial}{\partial t}\left(\varepsilon\, c_t\, x_P\right) = p\, v \tag{3}$$

Total molar balance for the gas

$$\mathrm{div}\mathbf{N}_t + \frac{\partial}{\partial t}\left(\varepsilon\, c_t\right) = \left(p - a\right) v \tag{4}$$

Molar balance for the solid reactant B

$$c_{B_0} \frac{\partial f_B}{\partial t} = -bv \tag{5}$$

Molar balance for the solid product Q

$$\frac{\partial c_Q}{\partial t} = qv \tag{6}$$

Molar balance for the inert solid J

$$c_J = c_{J_0} \tag{7}$$

Heat balance

$$c_{V_P} \frac{\partial T}{\partial t} + \text{div}(-\lambda_e \, \mathbf{grad} T) = v(-\Delta_r H) \tag{8}$$

This last balance supposes that the heat transported and stored by the gas is negligible and that the gas and the solid are at the same temperature at all points in the pellet.

2.2.2. Expressions for the molar flux densities

Calculation of the gaseous diffusion fluxes is a complex problem in the case of a multicomponent mixture (Taylor and Krishna, 1993), and even more complicated when the gas flows through the pores of a solid (Mason and Malinauskas, 1983). The most rigorous formulation consists in writing the Stefan-Maxwell equations which express the partial pressure gradients as a function of all the fluxes. However, the "inverse" form of these equations does not lend itself to a convenient resolution of the balance equations (eqs. 2 to 8). A simplified formulation is used here for the fluxes, involving the total flux and effective diffusion coefficients for each species. Thus,

$$\mathbf{N}_A = x_A \mathbf{N}_t - D_{A_e} c_t \, \mathbf{grad} x_A \tag{9}$$

and

$$\mathbf{N}_P = x_P \mathbf{N}_t - D_{P_e} c_t \, \mathbf{grad} x_P \tag{10}$$

The effective diffusion coefficients $D_{A_e}$ and $D_{P_e}$ are functions of the composition of the mixture ($x_A$, $x_P$ and $x_I$), the binary diffusivities ($\mathcal{D}_{AP}$, $\mathcal{D}_{AI}$ and $\mathcal{D}_{PI}$), the porous structure (porosity and tortuosity) and, possibly, the Knudsen diffusion coefficients. Different diffusion models, adapted to each problem, can supply these functions. An example is given in section 2.3.2.

It was also possible to write

$$\mathbf{N}_i = -D'_{i_e} c_t \, \mathbf{grad} x_i \qquad \text{where i=A or P} \tag{11}$$

with another type of effective diffusion coefficient, without referring to the total flux (Schneider, 1976). However, we preferred to introduce the total flux,

$$N_t = N_A + N_P + N_I \tag{12}$$

which in the present model results only from the non-equimolarity and anisothermality (eq. 4), in order to anticipate future generalisation to the non-isobaric case, in which a total flux would result from the viscous flow induced by a pressure gradient.

### 2.2.3. Other equations

The equations used to calculate the secondary variables $x_I$, $c_t$, $c_A$, $c_P$, $c_I$, $c_B$, $c_{V_p}$, and $X_B$ are listed below.

$$x_A + x_P + x_I = 1 \tag{13}$$

$$c_t = \frac{P}{RT} \tag{14}$$

$$c_i = x_i c_t \qquad \text{where i=A, P or I} \tag{15}$$

$$c_B = c_{B_0} f_B \tag{16}$$

$$c_{V_p} = c_B c_{p_B} + c_Q c_{p_Q} + c_J c_{p_J} \tag{17}$$

$$X_B = 1 - \frac{1}{V_p} \int_{V_p} f_B dV_p \tag{18}$$

### 2.2.4. Expression of the rate law

The heterogeneous gas-porous solid reaction takes place on part of the gas-solid surface. The surface reaction rate $\mathcal{R}_s$ is usually expressed as:

$$R_s = k_r \left( c_A^n - \frac{c_P^l}{K_{eq}} \right) \tag{19}$$

where $k_r$ is the rate constant, assumed to obey the Arrhenius law,

$$k_r = k_0 e^{-\frac{E_a}{RT}} \tag{20}$$

and n and l are the partial orders with respect to A and P. These orders are related to the stoichiometric coefficients by the equation

$$\frac{l}{p} = \frac{n}{a} \tag{21}$$

Note that the power law (eq. 19) is not appropriate for all types of reaction kinetics. For reaction kinetics which comply with the Langmuir-Hinshelwood model, the surface rate would be given rather by

$$R_s = k_{LH} \frac{c_A - \frac{c_P}{K_{eq}}}{1 + K_{LH,A} c_A + K_{LH,P} c_P} \quad (22)$$

This type of kinetic law is also provided for in the present model.

The reaction rate v included in the balance equations is a volume reaction rate and is related to the surface reaction rate via the specific reaction surface area. It is here assumed that a single chemical process controls the rate (no limitation by nucleation or diffusion processes) on the scale of the actual elementary grains of solid. It is then convenient to write

$$v = R_s a_0 s(f_B) \quad (23)$$

where $a_0$ is the initial specific area and where the function $s(f_B)$ reflects the variation of the reaction surface area as a function of the local conversion. This function depends on the structural model chosen:

- $s(f_B) = 1$ if the reaction surface area is constant;
- $s(f_B) = f_B^m$ generalizes the variation of the surface area corresponding to the grain models; this expression is used in the present model; the exponent m is often called, somewhat improperly, the "order" of the reaction with respect to the solid B; m is equal to $1/2$ for grains in the form of long cylinders and $2/3$ for spherical grains; m=1 has often been used in the pseudo-homogeneous models, but this value has no real physical significance;
- $s(f_B) = f_B \sqrt{1 - \psi \ln f_B}$ corresponds to the Random Pore Model of Bhatia and Perlmutter (1983), $\psi$ being a structure parameter.

The reaction rate v is therefore a function of the principal variables T, $x_A$, $x_P$, and $f_B$.

2.2.5. Initial and boundary conditions

At $t = 0$,

$$x_A(r,0) = x_{A_0}(r) \tag{24}$$

$$x_P(r,0) = x_{P_0}(r) \tag{25}$$

$$N_t(r,0) = 0 \tag{26}$$

$$T(r,0) = T_0(r) \tag{27}$$

$$f_B(r,0) = f_{B_0}(r) \tag{28}$$

These functions of the radius are usually simple constants.

At $t > 0$, at $r = 0$, because of the spherical symmetry,

$$\frac{\partial x_A}{\partial r} = \frac{\partial x_P}{\partial r} = N_t = \frac{\partial T}{\partial r} = 0 \tag{29}$$

at $r = R$, for matter, a convective exchange with the external gas is considered and, for heat, convective and radiative exchanges,

$$\left(D_{A_e}\frac{\partial x_A}{\partial r}\right)_R = k_{g_A}\left[x_{A_\infty} - x_A(R)\right] \tag{30}$$

$$\left(D_{P_e}\frac{\partial x_P}{\partial r}\right)_R = k_{g_P}\left[x_{P_\infty} - x_P(R)\right] \tag{31}$$

$$\left(\lambda_e\frac{\partial T}{\partial r}\right)_R = h\left[T_{g_\infty} - T(R)\right] + E\sigma\left[T_p^4 - T(R)^4\right] \tag{32}$$

The flux equalities (eqs. 30 and 31) concern the diffusion fluxes and not the total fluxes, as recommended by Bird et al. (1960). The emissivity factor E is a function of the emissivity of the external surface of the pellet and the emissivity of any solid wall to which the pellet surface radiates. Equations (30) to (32) generally vary with time.

*2.3. Expression of the parameters*

The physico-chemical and thermal parameters involved in the previous equations are often considered as constants in the literature. However, in the case of non-isothermal reactions, the variations with temperature and, where applicable, with composition, must be taken into account. It will therefore be considered that the specific heats, the enthalpy of reaction and its equilibrium constant are functions of temperature, and that they are known for a given reaction and constituents. The other parameters are determined as follows.

### 2.3.1. Transfer conductances

$k_{g_A}$, $k_{g_P}$ and h are calculated from classical correlations expressing the Sherwood or Nusselt numbers as a function of the Reynolds, and Schmidt or Prandtl numbers. For an isolated pellet in a gas flow, the Ranz and Marshall (1952) correlation gives

$$Sh_i = 2 + 0.6 Re^{\frac{1}{2}} Sc_i^{\frac{1}{3}} \quad \text{where i = A or P} \quad (33)$$

and

$$Nu = 2 + 0.6 Re^{\frac{1}{2}} Pr^{\frac{1}{3}} \quad (34)$$

with

$$Sh_i = \frac{2R k_{g_i}}{D_{if}} \quad Re = \frac{2R \rho_{gf} u_{g\infty}}{\mu_{gf}} \quad Sc_i = \frac{\mu_{gf}}{\rho_{gf} D_{if}} \quad Pr = \frac{c_{p_g f} \mu_{gf}}{\lambda_{gf}} \quad (35)$$

The parameters with the subscripts f are calculated for so-called "film" conditions, that is, for a mean composition and temperature between the values within the external bulk gas and the values at the external surface of the pellet. The viscosity of the gas mixture is calculated using the Hernig and Zipperer formula and its conductivity using the Friend and Adler one (Perry, 1963). Finally, the diffusivities $\mathcal{D}_{Af}$ and $\mathcal{D}_{Pf}$ of A and P in the gas mixture are determined as a function of the composition using the approach described below for the effective diffusivities.

### 2.3.2. Effective diffusivities

The influence of the porous texture of the pellet on the effective diffusivities is given by

$$D_{i_e} = \frac{\varepsilon}{\tau} D_{im} \quad (36)$$

where $\tau$ is the tortuosity coefficient and $D_{im}$ the pseudo-diffusion coefficient for the species i in the gas mixture within the pores.

The diffusion coefficients $D_{im}$ can be calculated from various models taking into account the diffusion regime (molecular, Knudsen) and the multicomponent nature of the gas. For example, the application presented in section 4 concerns the diffusion of three species in pores sufficiently large for Knudsen diffusion not to be limiting. The coefficients $D_{im}$ are then determined from the Stefan-Maxwell equations for a ternary mixture under isobaric conditions:

$$-c_t \, \mathbf{grad} \, x_i = \sum_{j=1}^{n} \frac{N_i x_j - N_j x_i}{D_{ij_e}} \quad \text{i=1: gas A; i=2: gas P; i=3: gas I} \quad (37)$$

- 12 -Assuming that

$$N_I = 0 \text{ (inert gas) and } p\, N_A = -a\, N_P \text{ (stoichiometry)} \tag{38}$$

as in pseudo-steady state conditions[†], gives

$$-c_t\, \mathbf{grad}\, x_A = N_A \left( \frac{\frac{p}{a} x_A + x_P}{D_{AP_e}} + \frac{x_I}{D_{AI_e}} \right) \tag{39}$$

and

$$-c_t\, \mathbf{grad}\, x_P = N_P \left( \frac{x_A + \frac{a}{p} x_P}{D_{AP_e}} + \frac{x_I}{D_{PI_e}} \right) \tag{40}$$

Comparing equations (9) and (39), and (10) and (40), finally yields

$$D_{A_e} = \frac{\varepsilon}{\tau} \frac{1 - \left(1 - \frac{p}{a}\right) x_A}{\frac{\frac{p}{a} x_A + x_P}{D_{AP}} + \frac{x_I}{D_{AI}}} \tag{41}$$

and

$$D_{P_e} = \frac{\varepsilon}{\tau} \frac{1 - \left(1 - \frac{a}{p}\right) x_P}{\frac{x_A + \frac{a}{p} x_P}{D_{AP}} + \frac{x_I}{D_{PI}}} \tag{42}$$

The binary diffusion coefficients $\mathcal{D}_{AP}$, $\mathcal{D}_{AI}$ and $\mathcal{D}_{PI}$ are calculated as a function of temperature using the Chapman-Enskog formula. Equations (41) and (42) are similar to those given by Eddings and Sohn (1993). However, to finish their calculation, these authors then assume a linear dependence of $D_{i_e}$ on $x_i$, which is not necessary in the present model.

### 2.3.3. Effective thermal conductivity

The effective thermal conductivity which appears in the heat balance (eq. 8) accounts for the conduction through the solid, the conduction through the gas, and the possible influence of gas flow on dispersive heat transport and of radiation at high temperature. It is thus a function of the temperature, the composition of solid and gas and the porous structure of the pellet. Despite this complexity, the effective thermal conductivity can be assessed from correlations of the literature, like these of Yagi and Kunii or Zehner and Schlünder (see Froment and Bischoff, 1990). It can also be measured but, since it is an effective and not an intrinsic parameter, measurements have to be made under conditions of temperature and gas composition close to those prevailing during the reaction.

---

[†] Hindmarsh and Johnson (1988) assert that the equality $p\, N_A = -a\, N_P$ remains almost exact in the general case, after a short initial transient period.

*2.4. Numerical method*

The partial differential equations (eqs. 2-8) are rendered discrete using the finite volume method (Patankar, 1980) with an implicit formulation. The implicit formulation has the advantage that the time increments required for its convergence are not too small. However, it is somewhat more tedious to program than the explicit formulation and requires convergence iterations at each time increment.

For the spatial discretisation, different ways of dividing the pellet volume were tested (equal volumes, equal radii, etc). Identical results were obtained, even for a fairly coarse mesh with only 20 cells. From 20 on, the number of cells has no influence on the results, whatever the reaction and the physical parameters. It was finally decided to divide the volume into 100 cells of equal radial thickness.

For the time discretisation, two options are provided in the program: a calculation at constant time step or a calculation at adaptive time steps. In the latter case, the time step is continuously increased to ensure a progression at constant increments of the degree of conversion. It is this option that is generally used, since it enables a considerable savings in computing time at the end of the reaction, when the conversion rate becomes slow. In general 100 or 200 conversion increments are used, up to complete conversion.

The flow chart for the calculation is shown in Fig. 1. The program code is written in Fortran.

The convergence iterations are commanded by a residue calculation involving all the principal variables. The number of iterations necessary varies depending on whether the reaction is taking place in isothermal or anisothermal conditions, and in the chemical or diffusional kinetic regimes. In an isothermal chemical regime, two or three iterations suffice, whereas up to 1500 are necessary in a non-isothermal diffusional regime. The total computing time to complete conversion varies in consequence. It is typically a few minutes of CPU on a Hewlett-Packard HP9000/755 workstation.

Exothermic reaction case

The treatment of exothermic reactions is difficult from the numerical viewpoint, due to their "autocatalytic" nature : the reaction liberates heat, leading to an increase in temperature, which accelerates the reaction rate, and so on. This characteristic rapidly causes divergence of an iterative calculation.

The source term, $s_{th}$, in the heat balance (eq. 8) can be expressed as a function of temperature considering the Arrhenius law for $k_r$ or $k_{LH}$

$$s_{th} = v(-\Delta_r H) = k_{th}\, e^{-\frac{E_a}{RT}} \tag{43}$$

where $k_{th}$ is itself a function of temperature via $\Delta_r H$, $c_t$, $K_{eq}$, etc., but whose variations are negligible compared to $e^{-\frac{E_a}{RT}}$, and for reasoning can therefore be considered to be constant. For the iterative calculation to be correct, the source term must be linearized in the form (Patankar, 1980)

$$s_{th} = s_0 + s_T T \quad \text{with} \quad s_0 \geq 0 \quad \text{and} \quad s_T \leq 0 \tag{44}$$

The ideal linearization is

$$s_0 = s^* - T^* \left(\frac{\partial s_{th}}{\partial T}\right)^* \quad \text{and} \quad s_T = \left(\frac{\partial s_{th}}{\partial T}\right)^* \tag{45}$$

where * identifies the values at the previous iteration. However, this linearization is obviously not applicable to $s_{th}$, which increases rapidly with temperature. The simple method which consists in taking

$$s_0 = s_{th}^* \quad \text{and} \quad s_T = 0 \tag{46}$$

is suitable for weakly exothermic reactions, but is insufficient for strongly exothermic reactions and/or those with a high activation energy. It should be noted in this respect that authors who have treated this problem numerically failed to simulate such reactions (Hindmarsh and Johnson, 1988).

After having unsuccessfully tested different possibilities, such as solution of the heat balance in terms of enthalpy and not temperature, or recourse to an explicit solution to avoid iterations, but with the need to use ridiculously short time steps, we developed an original method for treating the thermal source term. This method converges in all cases. We have called it "the two-temperature method".

The basic idea is to consider two temperature variables, one, $T_v$, being used in the exponential term of the Arrhenius law, and the other, T, everywhere else, and to gradually make these two temperatures converge. At the start of a new time increment, if the temperature increases, T is estimated by extrapolating the temperatures calculated at the preceding time increments

$$T = T_{t-\Delta t} + (T_{t-\Delta t} - T_{t-2\Delta t}) \quad (47)$$

and $T_v$ is deliberately overestimated by exaggerating the temperature rise by a factor $C_{T_v}$

$$T_v = T_{t-\Delta t} + C_{T_v}(T_{t-\Delta t} - T_{t-2\Delta t}) \quad (48)$$

For the first iteration, the estimated value of T is used in all the equations, except for $e^{-\frac{E_a}{RT_v}}$ which is calculated with the overestimated value of $T_v$. At the end of the iteration, during which T is recalculated from the heat balance, the two values are relaxed in the following manner

$$T_v = r_{T_v} T_v + (1 - r_{T_v}) T^* \quad (49)$$

$$T = r_T T^* + (1 - r_T) T \quad (50)$$

The iterations are then continued up to convergence, where $T_v = T^* = T$. The process is illustrated in Fig. 2.

In fact, this method amounts to relaxing the temperature and relaxing non-linearly an artificially decreasing source term. To ensure convergence when ($-\Delta_r H$) and $E_a$ are high, it is sufficient to increase $C_{T_v}$ and to enhance the under-relaxation. For the simulations in section 4, it was considered that $C_{T_v} = 20$, $r_{T_v} = 0.5$ and $r_T = 0.96$.

*2.5. Results of the model*

Given a reaction of the (eq. 1) type, the model calculates all the variables and parameters described above. The results are either visualized dynamically, i.e. during the calculation, in the form of isovalue maps $f_B(r)$, $T(r)$, $x_A(r)$, etc., or edited at the end of the calculation as files containing the values $f_B(r)$, $T(r)$, $x_A(r)$, etc. at different times or different degrees of conversion chosen in advance, together with the values $f_B(t)$, $T(t)$, $x_A(t)$, etc. at different preselected radii.

## 3. VALIDATION OF THE MODEL

In order to validate the model, the results obtained were compared with different analytical and numerical solutions published in the literature.

*3.1. Isothermal case*

3.1.1. Comparison with Ishida and Wen's model (1968a)

Ishida and Wen developed an isothermal pseudo-steady state model for an equimolar gas-solid reaction in a spherical porous pellet, without any inert species. The reaction was assumed to be of first order with respect to the reactive gas and to be independent of the concentration of reactive solid (m=0). Their model distinguishes two periods: a period I during which the reaction is terminated nowhere in the pellet and a period II, during which an external zone where the reaction is completed coexists with an internal zone where the reaction continues. One of the advantages of this model is the possibility of studying different kinetic regimes with an analytical, i.e. exact, solution.

Figure 3 compares the variation of the degree of conversion as a function of a reduced time, calculated either with the equations of Ishida and Wen, or with the present model. The reduced time is defined as $t/t_{ch}$ where $t_{ch}$ is the characteristic time for the chemical reaction, that is, the time that would be necessary to obtain complete conversion if the reaction occurred in the chemical regime. In the case exemplified, the regime is mixed, so that the diffusional and chemical resistances are of the same order of magnitude, as indicated by the modulus $\phi = 1$, with

$$\phi^2 = \frac{\text{resistance due to internal diffusion}}{\text{resistance due to chemical reaction}} = \frac{t_{dif}}{t_{ch}} \tag{51}$$

where $t_{dif}$ is the characteristic diffusion time, corresponding to the time necessary to obtain complete conversion in the diffusional regime.

The modified Sherwood number Sh' represents the efficiency of external mass transfer

$$Sh' = \frac{k_{g_A} R}{D_{A_e}} \tag{52}$$

The agreement between the curves calculated by the two methods is perfect.

3.1.2. Comparison with the Grain Model of Szekely et al. (1976)

The Grain Model presented in the introduction constitutes an indispensable reference for any simulation of gas-solid reaction kinetics. Except for the assumption of sphericity, the present model is more general than the Grain Model and can therefore predict all its results. The only case with which the present model cannot deal directly is that where a diffusional resistance exists in the layer of solid product formed around the cores of the small grains. This situation was treated as an extension of the basic Grain Model by Szekely et al. (1976).

Figure 4 shows the variation with time of the degree of conversion calculated using both the Grain Model and the present model, for the cases of a mixed regime ($\phi_s = 1$) and a mixed but diffusion-dominated regime ($\phi_s = 2$), with no external resistance (Sh' $\rightarrow \infty$). As could be expected, the two models give identical results.

*3.2. Anisothermal case*

The strong point of the present model is its ability to treat the interaction among reaction, diffusion and heat transfer in the case of endo- and exothermic reactions. As pointed out in the introduction, few similar calculations are described in the literature, except under pseudo-steady state conditions, whereas the heat balance tends to be incompatible with this assumption, owing to the thermal inertia of the solid.

The most appropriate work for a direct validation of the thermal part of the present model is that of Luss and Amundson (1969). These authors elegantly treated the case of a diffusion-controlled exothermic reaction and obtained an analytical solution for calculating the temperature at any time and at any point in the pellet. The principal limitation of their calculation is the use of constant physical parameters, whereas the mass diffusivity and the thermal conductivity vary with temperature.

The temperature profiles calculated by Luss and Amundson are compared in Fig. 5 with those obtained with the present model, for the conditions detailed in Luss and Amundson's original paper. The agreement between the two calculations is very good, validating the numerical code in this anisothermal case.

The variable $\xi$ on the abscissa of Fig. 5 is r/R, while the ordinate $\theta$ is the dimensionless temperature rise, defined by

$$\theta = \frac{hR(T - T_\infty)}{6(-\Delta_r H) D_{A_e} c_{A_\infty}} \qquad (53)$$

The parameter y represents $r_c/R$ where $r_c$ is the position of the reaction front. Actually, under diffusion control, on the scale of the pellet, the reaction is localised along a front separating the solid reactant B, on the inside, from the solid product Q, on the outside. The position of the front corresponds to the angular point on each of the curves. The heat of reaction is liberated at this location. Thus, at the start of the reaction (y = 0.96 to 0.88), the temperature maximum is situated near the surface, while the core is gradually heated by conduction until the temperature becomes uniform from the centre to the front. Subsequently, the outer layers are cooled by external exchange, while the maximum temperature, at the centre, decreases more slowly. It should be noted that steady-state regime models are incapable of predicting this type of behaviour, since they always place the maximum temperature at the centre of the pellet.

## 4. APPLICATION TO THE OXIDATION OF ZINC SULPHIDE

After having shown the numerical validity of our model by the comparisons described above, we used it to simulate a highly exothermic gas-solid reaction, the oxidation of zinc sulphide

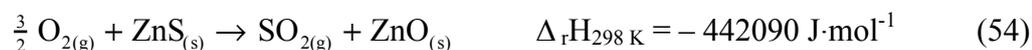

$$\tfrac{3}{2} O_{2(g)} + ZnS_{(s)} \rightarrow SO_{2(g)} + ZnO_{(s)} \qquad \Delta_r H_{298\,K} = -442090 \text{ J·mol}^{-1} \qquad (54)$$

We studied the kinetics of this reaction by thermogravimetry, using synthetic porous pellets of pure ZnS and various $O_2+N_2$ mixtures. The experiments, as described by Galant François (1995), associated gravimetric measurements and thermal measurements which consisted in recording the temperature at the pellet centre. In certain runs, the measured temperature rise was considerable: more than 500 °C. To our knowledge, such a marked thermal effect had never previously been studied in the field of gas-solid reactions, either experimentally or theoretically. The simulation of these experiments therefore constituted an excellent test of the model.

The model enabled us to interpret the experimental results quantitatively. We showed (Patisson et al., 1996) that (i) owing to the rise in temperature due to the reaction, the kinetic regime was always diffusion controlled, even when the external temperature conditions would suggest a chemical or mixed regime; (ii) the external resistance to mass transfer was not negligible, even for high gas flowrates; (iii) the reacting pellet showed strong temperature gradients, localised in the shell layer of oxide.

All the experiments were successfully simulated. We only report here results exemplifying two different situations: an experiment under air at 850 °C and another under pure oxygen at 550 °C. The data used for the simulation were detailed by Galant François (1995). The main parameters are also given in Table 1. Figure 6 compares the temperatures and degrees of conversion measured as a function of time with those calculated using the model, for both experiments. The temperature was measured at the centre of the pellet (r = 0), and calculated at the centre (r = 0), at an intermediate radius (r = 3 mm), and at the external surface (r = 5 mm). The agreement between the measured and calculated values is quite satisfactory and provides a further validation of the model, this time experimental.

Table 1.  Parameters used for the simulation of the oxidation of a zinc sulphide pellet

| | |
|---|---|
| pellet | initially pure ZnS; R = 5 mm; $c_{B_0}$ = 12810 to 16550 mol·m$^{-3}$ according to the sample; $a_0 = 9.6 \times 10^5$ m$^{-1}$; $\varepsilon = 0.52$; $\tau = 1.4$ |
| reaction rate | $R_s = 1.421 \times 10^{11} \exp\left(-\dfrac{25250}{RT}\right) c_{O_2}^{\frac{2}{3}}$ mol·m$^{-2}$·s$^{-1}$; $m = \dfrac{2}{3}$ |
| external conditions | $u_{g_\infty} = 5.17 \times 10^{-2}$ m·s$^{-1}$ at 0 °C and 1 atm; $T_p = T_{g_\infty} - 50$ K from measurements |
| heat transfer | $E_{ZnO} = 0.6$ adjusted; $\lambda_e = 0.2 + 2.5 \times 10^{-10}$ T$^3$ W·m$^{-1}$·K$^{-1}$ adjusted from a value $\lambda_e = 0.21$ measured at 25 °C for a ZnO pellet |

## 5. CONCLUSIONS

A kinetic model for gas-solid reactions was developed which is sufficiently general to enable simulation in the transient regime of reactions which may be exo- or endothermic, equimolar or not, reversible or irreversible, with the possible presence of inert gases and solids. The model was validated by comparing its predictions with analytical and numerical solutions published in the literature, and was applied to the exothermic reaction involved in the oxidation of zinc sulphide.

In the present paper, it has been chiefly endeavoured to describe the anisothermal specificity of the model, rather than to give an exhaustive presentation of all its potentialities. In particular, future articles will show the advantages of its ability to deal with transient regime and non-equimolarity. It is also planned to generalize the model to cases where the porosity varies with the degree of conversion.

**NOTATION**

| | |
|---|---|
| A | gaseous reactant |
| a, b | stoichiometric coefficients |
| $a_0$ | initial specific surface area of the pellet, $m^2 \cdot m^{-3}_{pellet}$ |
| B | solid reactant |
| $C_{T_v}$ | constant in eq. (48) |
| $c_A, c_P$ | molar concentration of species A, P in the gas, $mol \cdot m^{-3}$ |
| $c_{B_0}, c_{J_0}$ | initial molar concentration of species B, J in the pellet, $mol \cdot m^{-3}_{pellet}$ |
| $c_B, c_J, c_Q$ | molar concentration of species B, J, Q in the pellet, $mol \cdot m^{-3}_{pellet}$ |
| $c_{p_B}, c_{p_J}, c_{p_Q}$ | specific heat of species B, J, Q, $J \cdot mol^{-1} \cdot K^{-1}$ |
| $c_{p_g}$ | specific heat of the gas, $J \cdot kg^{-1} \cdot K^{-1}$ |
| $c_t$ | overall molar concentration of the gas, $mol \cdot m^{-3}$ |
| $c_{V_p}$ | specific heat of the pellet, $J \cdot m^{-3} \cdot K^{-1}$ |
| $D_{A_e}, D_{P_e}$ | effective diffusivity of species A, P in the pellet, $m^2 \cdot s^{-1}$ |
| $\mathcal{D}_{AP}, \mathcal{D}_{AI}, \mathcal{D}_{PI}$ | binary diffusivity of pairs A-P, A-I and P-I, $m^2 \cdot s^{-1}$ |
| $\mathcal{D}_{Af}, \mathcal{D}_{Pf}$ | diffusivity of A, P in the gaseous mixture under film conditions, $m^2 \cdot s^{-1}$ |
| $D_{im}$ | pseudo-binary coefficient of diffusion of species i in the gaseous mixture, $m^2 \cdot s^{-1}$ |
| E | emissivity factor |
| $E_a$ | activation energy, $J \cdot mol^{-1}$ |
| $f_B$ | local fraction of B unreacted |
| h | external convective heat transfer coefficient, $W \cdot m^{-2} \cdot K^{-1}$ |
| I | inert gas |
| J | inert solid |
| $K_{eq}$ | equilibrium constant, $mol^{1-n} \cdot m^{-3(1-n)}$ |
| $k_0$ | frequency factor, $mol^{1-n} \cdot s^{-1} \cdot m^{2-3n}$ |
| $k_{g_A}, k_{g_P}$ | external mass transfer coefficient for species A, P, $m \cdot s^{-1}$ |
| $k_r$ | reaction rate constant, $mol^{1-n} \cdot s^{-1} \cdot m^{2-3n}$ |
| $k_{LH}, K_{LH,A}, K_{LH,P}$ | constants in eq. (22) |
| $k_{th}$ | pre-exponential factor in eq. (43), $W \cdot m^{-3}$ |
| $l$ | reaction order with respect to species P |
| $m$ | exponent of the reaction rate dependence on the fraction of solid B |
| $N_A, N_I, N_P$ | molar flux density of species A, I, P, $mol \cdot s^{-1} \cdot m^{-2}_{pellet}$ |
| $N_t$ | overall molar gas flux density, $mol \cdot s^{-1} \cdot m^{-2}_{pellet}$ |
| $n$ | reaction order with respect to species A |
| P | gaseous product |
| P | overall gas pressure, Pa |
| $p, q$ | stoichiometric coefficients |
| Q | solid product |
| R | radius of the pellet, m |
| R | ideal gas constant, $J \cdot mol^{-1} \cdot K^{-1}$ |
| $\mathcal{R}_s$ | rate of the chemical reaction per unit reaction surface area, $mol \cdot s^{-1} \cdot m^{-2}$ |
| $r$ | radial position in the pellet, m |
| $r_c$ | radial position of the reaction front in the pellet, m |
| $r_{T_v}, r_T$ | relaxation factors in eqs. (49) and (50) |
| $s(f_B)$ | function describing the change of reaction surface area with conversion |
| $s_0, s_T$ | source term factors in eq. (44), $W \cdot m^{-3}, W \cdot K^{-1} \cdot m^{-3}$ |
| $s_{th}$ | source term in the thermal balance, $W \cdot m^{-3}$ |
| T | temperature, K |
| $T_v$ | temperature in $e^{-\frac{E_a}{R T_v}}$, K |
| $t$ | time, s |
| $t_{ch}$ | characteristic time for chemical reaction, s |
| $t_{dif}$ | characteristic time for internal diffusion, s |
| $u_{g\infty}$ | velocity of the external gas, m·s$^{-1}$ |
| $V_p$ | pellet volume, m$^3$ |
| $v$ | rate of the chemical reaction per unit pellet volume, mol·s$^{-1}$·m$^{-3}_{pellet}$ |
| $X_B$ | overall degree of conversion |
| $x_A, x_I, x_P$ | molar fraction of species A, I, P in the gas |

*Greek letters*

| | |
|---|---|
| $\Delta_r H$ | heat of the reaction, J·mol$^{-1}$ |
| $\Delta t$ | time step, s |
| $\varepsilon$ | porosity of the pellet |
| $\lambda_e$ | effective thermal conductivity of the pellet, W·m$^{-1}$·K$^{-1}$ |
| $\lambda_g$ | thermal conductivity of the gas, W·m$^{-1}$·K$^{-1}$ |
| $\mu_g$ | viscosity of the gas, kg·m$^{-1}$·s$^{-1}$ |
| $\rho_g$ | density of the gas, kg·m$^{-3}$ |
| $\phi$ | reaction-diffusion modulus |
| $\sigma$ | Stefan constant, W·m$^{-2}$·K$^{-4}$ |
| $\theta$ | dimensionless temperature rise |
| $\tau$ | tortuosity coefficient of the pellet |
| $\xi$ | dimensionless radial position in the pellet |
| $\psi$ | structural parameter of the Random Pore Model |

*Dimensionless numbers*

| | |
|---|---|
| Nu | Nusselt number |
| Pr | Prandtl number |
| Re | Reynolds number |
| Sc | Schmidt number |
| Sh, Sh' | Sherwood number, modified Sherwood number |

*Subscripts*

| | |
|---|---|
| 0 | initial conditions |
| f | film conditions |
| i, j | species |
| ∞ | external gas conditions |

*Superscripts*

| | |
|---|---|
| * | previous iteration |

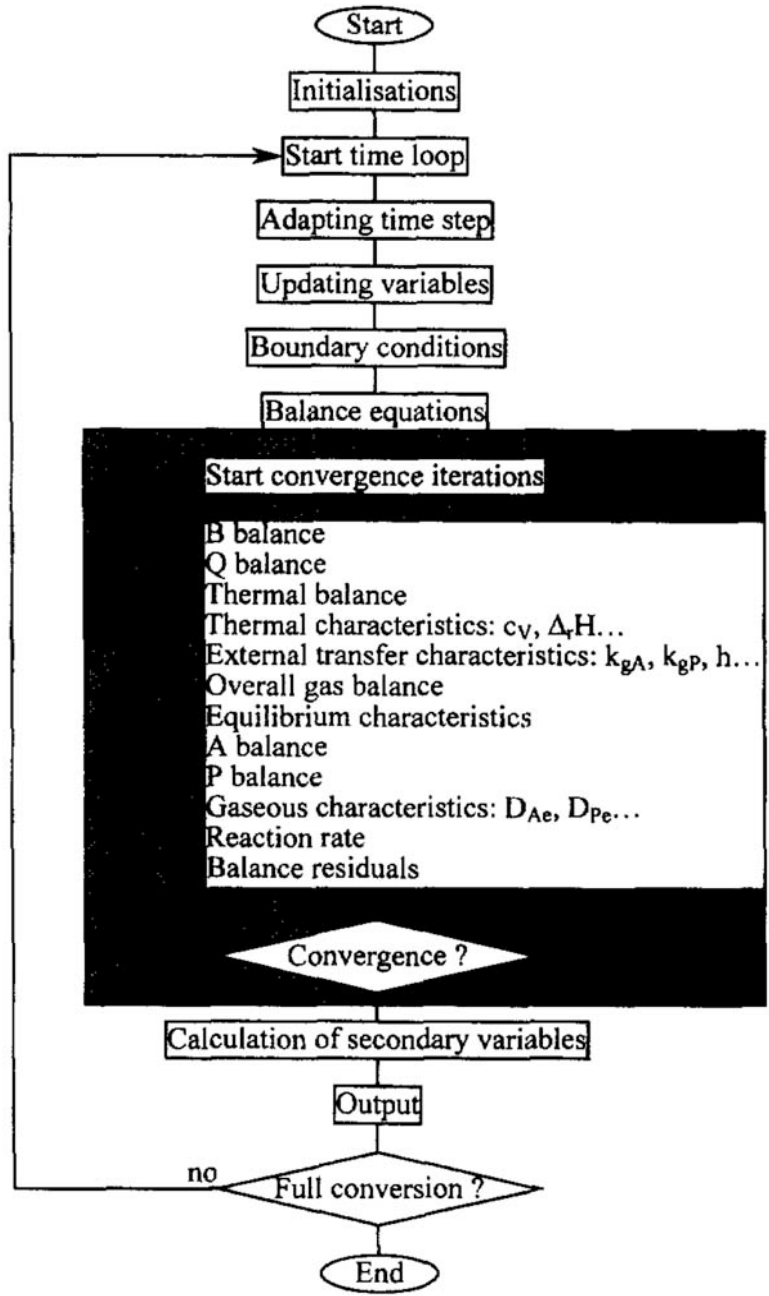

Fig. 1. Calculation flow chart.

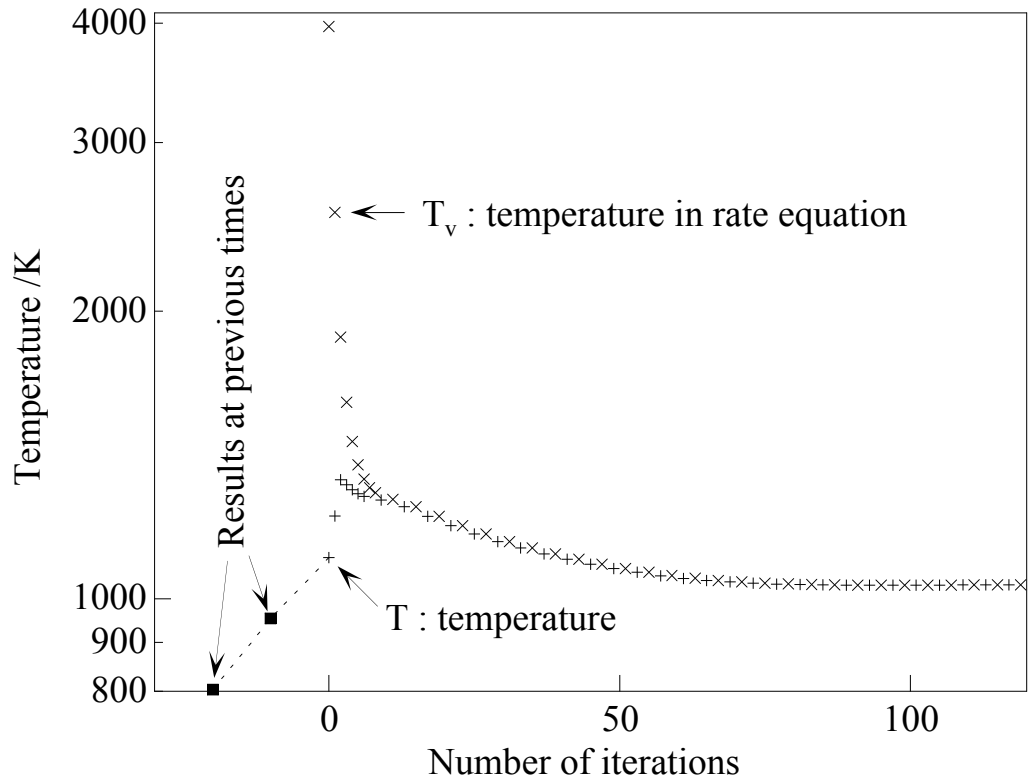

Fig. 2.  Typical evolution of temperature during
the iterative convergence process for an exothermic reaction

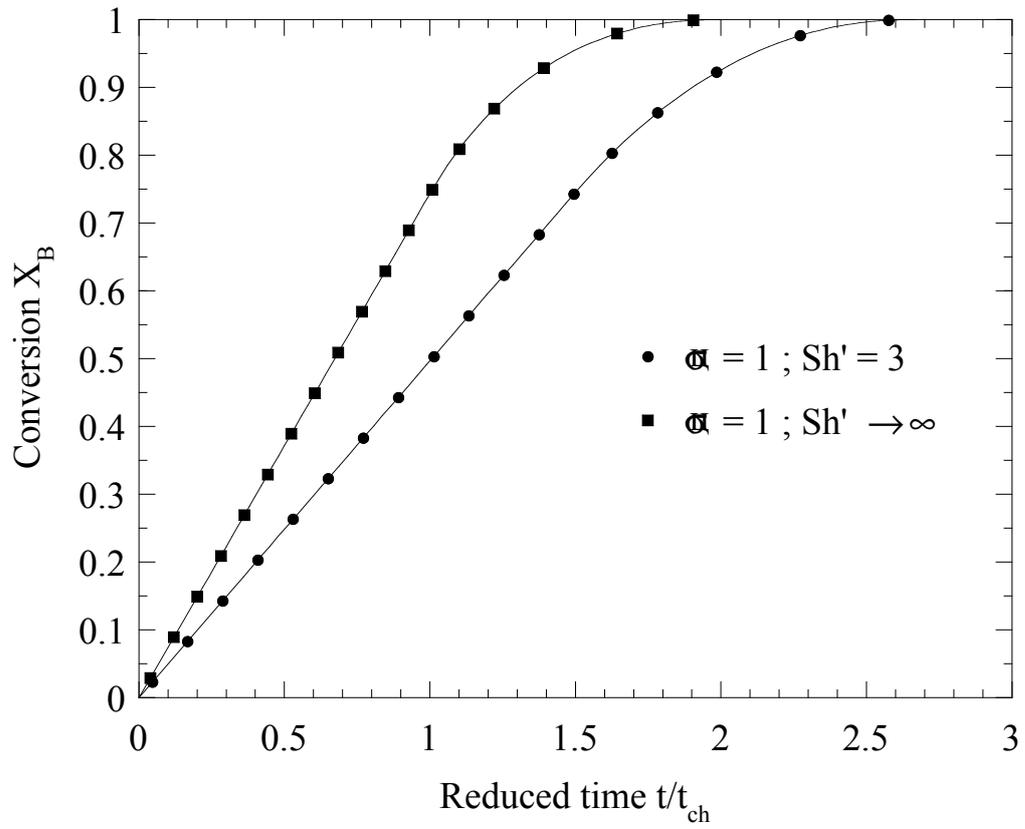

Fig. 3. Comparison of conversion vs time curves; analytical solution by Ishida and Wen (1968a) ——— , and present numerical solution ● ■

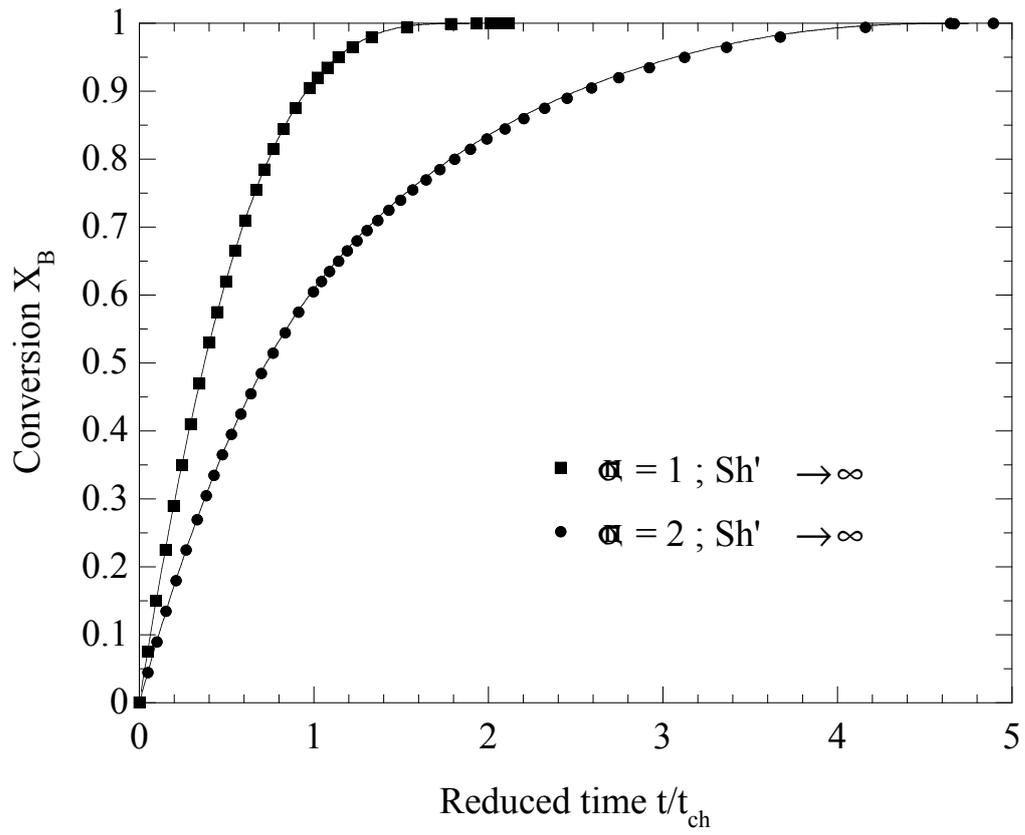

Fig. 4. Comparison of conversion vs time curves; numerical solutions by Szekely et al. (1976) ———, and present one ● ■

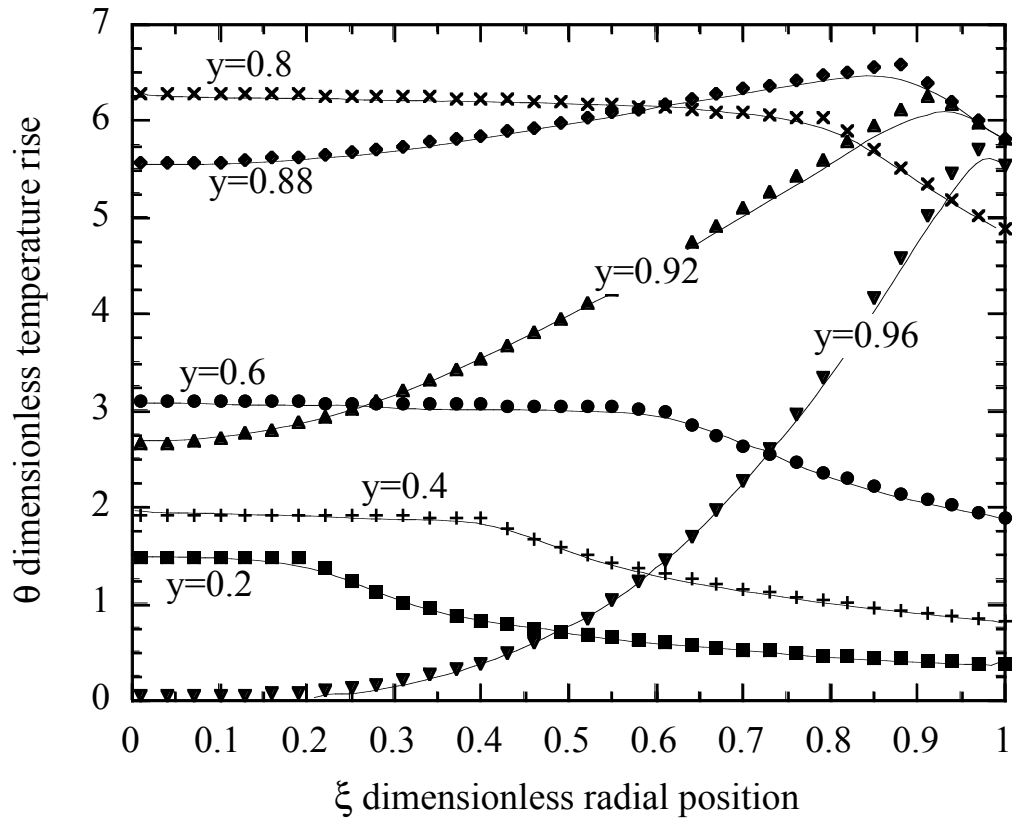

Fig. 5. Comparison of temperature profiles in a spherical pellet undergoing an exothermic reaction under diffusion control; analytical solution by Luss and Amundson (1969) ———, and present numerical solution ▼ ▲ ♦ × ● + ■

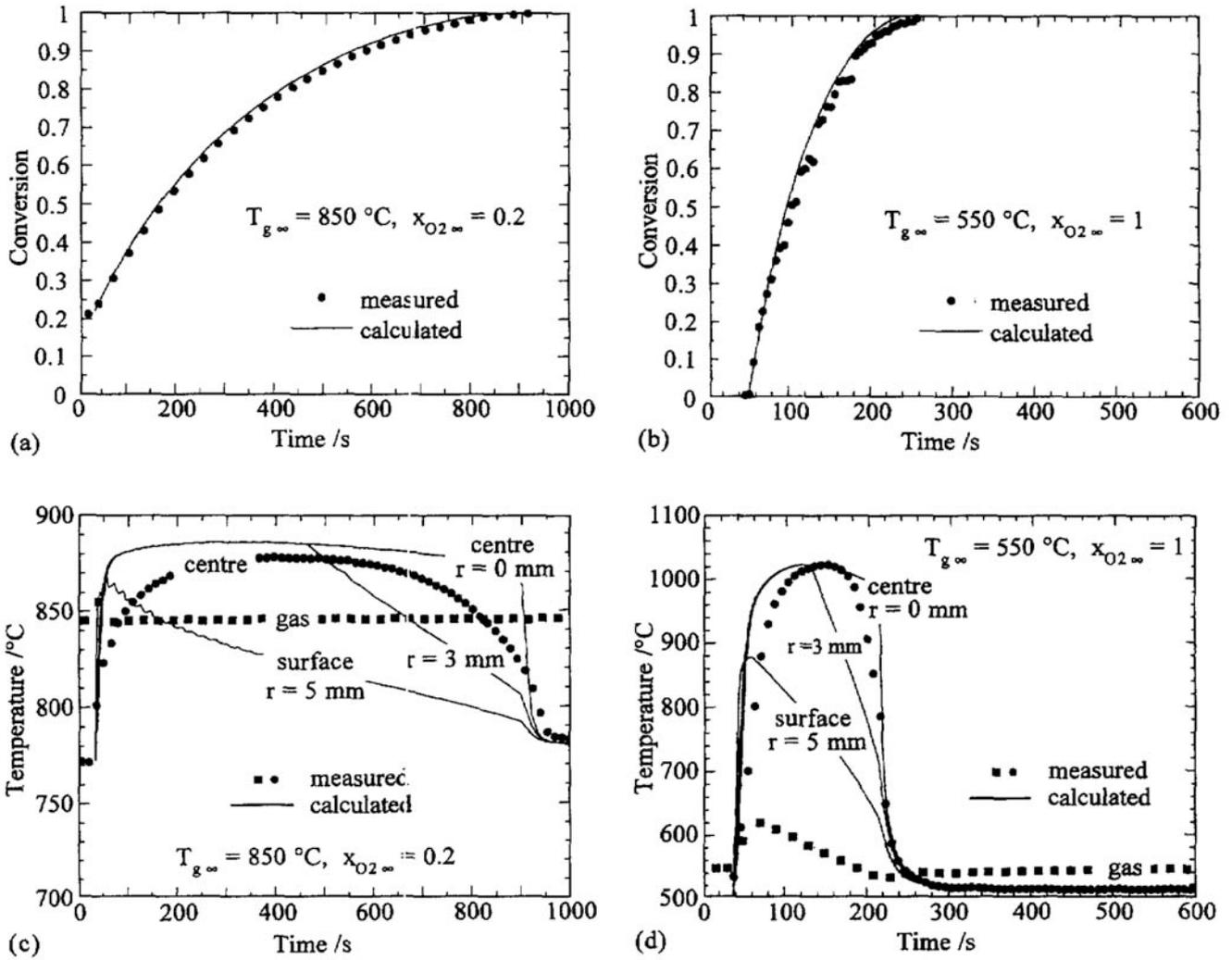

Fig. 6. Comparison of measurements and calculations from the model for the oxidation of a zinc sulphide pellet; (a) and (b): conversion vs time; (c) and (d): temperature vs time; (a) and (c): $T_{g\infty} = 850°C$ and $x_{O_2\infty} = 0.2$; (b) and (d): $T_{g\infty} = 550°C$ and $x_{O_2\infty} = 1$.